\documentclass[preprintnumbers,aps,pra,groupedaddress,twocolumn]{revtex4}

\usepackage{amsmath}
\usepackage{bm}
\usepackage{color}
\usepackage{hyperref}
\usepackage{graphicx}
\usepackage{amsfonts}

\hypersetup{
	colorlinks=true,
	linkcolor=blue,
	filecolor=blue,      
	urlcolor=blue,
	citecolor=blue,
}

\begin{document}

\title{Exceptional phase transition in a single Kerr-cat qubit}

\author{Pei-Rong Han$^{1}$}
\author{Tian-Le Yang$^{2,3}$}
\author{Wen Ning$^{2,3}$}
\author{Hao-Long Zhang$^{2,3}$}
\author{Huifang Kang$^{1}$}
\author{Huiye Qiu$^{1}$}
\thanks{E-mail: qiuhuiye@lyun.edu.cn}
\author{Zhen-Biao Yang$^{2,3}$}
\thanks{E-mail: zbyang@fzu.edu.cn}
\address{$^1$School of Physics and Mechanical and Electrical Engineering, Longyan University, Longyan 364012, China\\
$^2$Fujian Key Laboratory of Quantum Information and Quantum Optics, College of Physics and
Information Engineering, Fuzhou University, Fuzhou, Fujian 350108, China\\
$^3$Department of Physics, Fuzhou University, Fuzhou 350108, China}


\begin{abstract}

Exceptional points in non-Hermitian quantum systems give rise to novel genuine quantum phenomena. Recent explorations of exceptional-point-induced quantum phase transitions have extended from discrete-variable to continuous-variable-encoded quantum systems. However, quantum phase transitions driven by Liouvillian exceptional points (LEPs) in continuous-variable platforms remain largely unexplored. Here, we construct and investigate a Liouvillian exceptional structure based on a driven-dissipative Kerr-cat qubit. Through numerical simulations, we reveal a quantum phase transition occurring at the LEP characterized by a sudden change in dynamical behavior from underdamped oscillations to overdamped relaxations as visualized via Wigner functions and Bloch sphere trajectories. Notably the negativity of the Wigner function serves as a direct signature of genuine quantum coherence unattainable in conventional single-qubit non-Hermitian systems. 
Furthermore, we introduce the phase difference between the off-diagonal elements of the Liouvillian eigenmatrices as a novel parameter to quantify the transition.
Our results establish the Kerr-cat qubit as a novel continuous-variable setting for exploring dissipative quantum criticality and intrinsic non-Hermitian physics.

\end{abstract}
\maketitle

Non-Hermitian physics has attracted considerable attention in recent years, with research advancing from classical \cite{1,2,3,4,5,6,7,8,9,10,11,12,13,14} or semiclassical \cite{15,16,17,18,19,20,21,22,23,24,25,26,27,28,29,30} regimes into the full quantum domain \cite{31,32,33,34,35,36,37,38,39,40,41,42,43,44,45}. Recently, these studies have been extended from discrete-variable to continuous-variable quantum systems \cite{46,47}. These works explore distinct physical phenomena associated with Hamiltonian exceptional points (HEPs) and Liouvillian exceptional points (LEPs), such as entanglement phase transitions \cite{46} and non-Hermitian topology \cite{47}. An HEP is a singular point where both eigenvalues and eigenvectors of a non-Hermitian Hamiltonian coalesce; its experimental observation typically relies on post-selection \cite{27,28,29,30,31,32,33}. In contrast, an LEP refers to the coalescence of eigenvalues and eigenvectors of a Liouvillian superoperator, which arises without post-selection and inherently incorporates all quantum jumps—a hallmark of genuinely open quantum dynamics \cite{36,37,38,39,40,41,42,43,44,45}. While HEPs and LEPs share similar properties in the semiclassical limit, they exhibit fundamentally different behaviors in the quantum regime \cite{35}. However, quantum phase transitions driven by LEPs in continuous-variable systems remain largely unexplored.

In this paper, we construct the Liouvillian superoperator for a continuous-variable system based on a single driven-dissipative Kerr-cat qubit \cite{48,49,50,51,52,53,54,55}. Within this framework, the computational basis (Z-basis) is encoded using parity-symmetric cat states: the even-parity state $|\mathcal{C}_{\alpha}^{+}\rangle = \mathcal{N}_{\alpha}^{+}(|\alpha\rangle + |-\alpha\rangle)$ and the odd-parity state $|\mathcal{C}_{\alpha}^{-}\rangle = \mathcal{N}_{\alpha}^{-}(|\alpha\rangle - |-\alpha\rangle)$, where $|\pm\alpha\rangle$ are coherent states and $\mathcal{N}_{\alpha}^{\pm}=1/\sqrt{2(1\pm e^{-2|\alpha|^2})}$ are normalization constants. 
Accordingly, the X-basis states are defined as $|\pm X\rangle=|\pm\alpha\rangle$ and the Y-basis states as $|\pm Y\rangle=|\mathcal{C}_{\alpha}^{\mp i}\rangle=(|\mathcal{C}_{\alpha}^{+}\rangle\pm i|\mathcal{C}_{\alpha}^{-}\rangle)/\sqrt{2}$.
This encoding leverages the degenerate eigenstate manifold of a Kerr-nonlinear resonator under two-photon driving, which provides a naturally noise-biased qubit subspace \cite{48,55,58}. 

\begin{figure}[hbtp]
	\centering
	\includegraphics[width=3.5in]{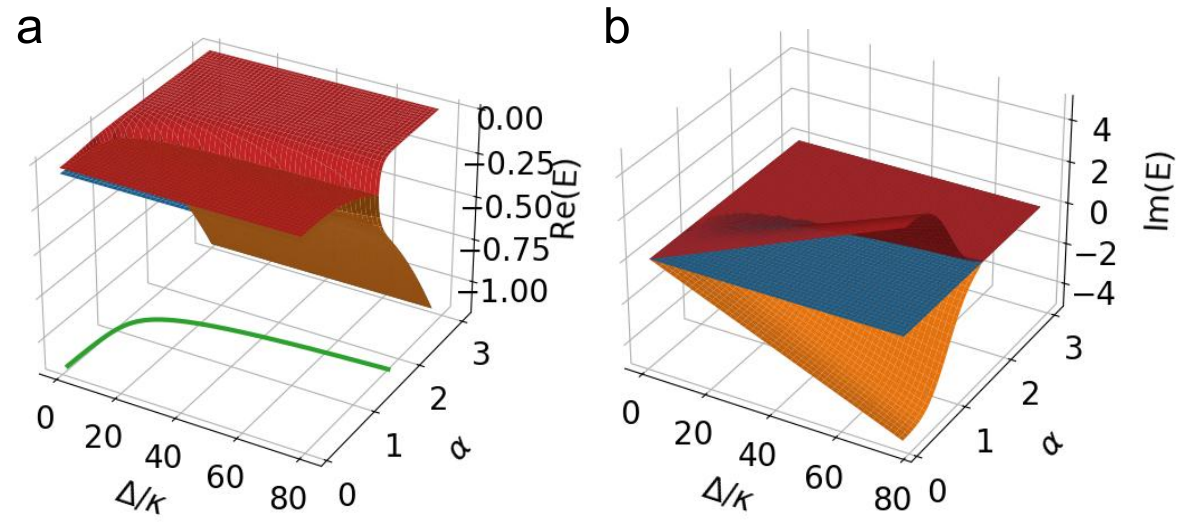}
	\caption{Eigenspectrum of the Liouvillian matrix $\mathcal{L}_{\mathrm{matrix}}$. Panels (a) and (b) depict the real and imaginary parts, respectively, with colors denoting different eigenvalues (excluding the steady-state eigenvalue $E_1=0$). The solid green curve in (a) indicates LEP2s in the $\Delta$–$\alpha$ parameter space at fixed single-photon loss $\kappa$.}
    \label{Fig1}
\end{figure}

The non-Hermitian character of the system's Liouvillian originates from single-photon loss. This dissipation channel mediates bidirectional quantum jumps between the even- and odd-parity cat states, analogous to the jumps described by a $\sigma_x$ operator in a conventional qubit. Introducing a detuning $\Delta$ between the two-photon drive frequency and the frequency of Kerr resonator lifts the degeneracy of the cat states, effectively implementing coherent rotations around the logical Z-axis of the qubit. By examining the evolution of the quantum state across different detunings, one can observe distinct dynamical behaviors in the vicinity of LEPs. Specifically, the system exhibits a phase transition induced by an LEP.

\begin{figure}[hbtp]
	\centering
	\includegraphics[width=3.4in]{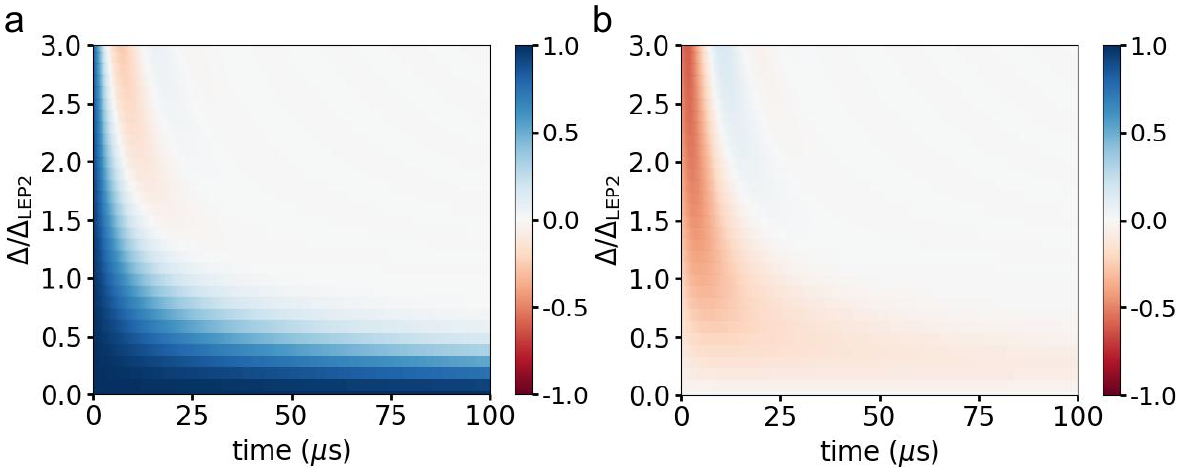}
	\caption{
    Time evolution of (a) $\langle X\rangle$ and (b) $\langle Y\rangle$ under different detunings $\Delta$.
    For $|\Delta| > \Delta_{\text{LEP2}}$, damped oscillations with exponential envelopes are observed.
    When $|\Delta| < \Delta_{\text{LEP2}}$, the system is overdamped without oscillation.
    At the critical point $|\Delta| = \Delta_{\text{LEP2}}$, critical damping occurs, leading to the fastest convergence to the steady state.
    The parameters are $\kappa/2\pi=10$ kHz, $K/2\pi=6.7$ MHz, $P/2\pi=15.5$ MHz and $\kappa_{\phi}=0$.}
	\label{Fig2}
\end{figure}

Our scheme is based on  a Kerr-nonlinear resonator with a two-photon driving. In the rotating frame, the Hamiltonian for the system is given by (we set $\hbar \equiv 1$ throughout this paper)
\begin{equation}
    H = \Delta a^{\dagger} a + Ka^{\dagger2} a^2 + P\left(a^{\dagger2}+a^2 \right),
\end{equation}
where $\Delta$ is the driving-resonator detuning, $K$ is the Kerr coefficient and $P$ is the amplitude of the two-photon driving. $a^{\dagger}$($a$) is the creation (annihilation) operator for the Kerr-nonlinear resonator. 
When $\Delta = 0$, the Hamiltonian admits both coherent states $|\pm\alpha\rangle$ and cat states
$|\mathcal{C}_{\alpha}^{\pm}\rangle$
as eigenstates for large $\alpha$, where $\alpha = \sqrt{P/K}$. Throughout this paper, we assume $K$ and $P$ are positive and real, hence $\alpha$ is also real. These cat states exhibit exact orthogonality
$\langle\mathcal{C}_{\alpha}^{k}|\mathcal{C}_{\alpha}^{j}\rangle = \delta_{kj}$, while the coherent states are merely
quasi-orthogonal for large $\alpha$ with $|\langle\alpha|{-}\alpha\rangle| = e^{-2\alpha^2}$. This coherent-state non-orthogonality
modifies exceptional points within the cat-state encoding subspace, quantified by the ratio
$p =\mathcal{N}_{\alpha}^{+}/\mathcal{N}_{\alpha}^{-}$.

In experiments, systems inevitably interact with their environments through single-photon loss, described by the annihilation operator $a$ with rate $\kappa$, and pure dephasing, described by the photon-number operator $a^\dagger a$ with rate $\kappa_\phi$.
The dynamics of this driven-dissipative system is governed by the Lindblad master equation
\begin{equation}
    \dot{\rho} \equiv \mathcal{L}(\rho) = -i [H,\rho] + \sum_{\mu}\mathcal{D}\left[\mathcal{O}_{\mu}\right]\rho,
    \label{master eq}
\end{equation}
where $\mathcal{L}$ is the Liouvillian superoperator, $\mathcal{D}\left[\mathcal{O}_{\mu}\right]$ are the dissipators associated with the jump operators $\mathcal{O}_{\mu}$, $\mathcal{D}[\mathcal{O}_{\mu}]\rho=\mathcal{O}_{\mu}\rho\mathcal{O}_{\mu}^{\dagger}-\frac{1}{2}\mathcal{O}_{\mu}^{\dagger}\mathcal{O}_{\mu}\rho-\frac{1}{2}\rho\mathcal{O}_{\mu}^{\dagger}\mathcal{O}_{\mu}$ and $\rho$ is the density matrix of the nonlinear resonator. 
\textcolor{black}{The model described here fits within the topological classification framework for driven-dissipative systems, where steady-state structures and fluctuation dynamics are characterized by graph invariants \cite{56,57}.}
While single-photon loss ($\mathcal{O}_1=\sqrt{\kappa}a$) typically introduces bit-flip errors that degrade the encoding subspace, it paradoxically drives the system toward steady states comprising statistical mixtures of parity-opposite cat states,
$\rho_{ss} = p_1|\mathcal{C}_{\alpha}^{\pm}\rangle\langle\mathcal{C}_{\alpha}^{\pm}|+p_2|\mathcal{C}_{\alpha}^{\mp}\rangle\langle\mathcal{C}_{\alpha}^{\mp}|$ ($p_1+p_2\simeq1$). 
In contrast, dephasing noise ($\mathcal{O}_2=\sqrt{\kappa_\phi}a^\dagger a$) remains suppressed when the environmental noise bandwidth is narrower than the system's energy gap \cite{52}, yet still modifies the Liouvillian spectrum and dynamical behavior.

To study the Liouvillian spectra and LEPs, we use a vectorized representation to obtain the matrix form of the Liouvillian superoperator. Conditioned on the relations $a|\mathcal{C}_{\alpha}^{\pm}\rangle=\alpha p^{\pm 1}|\mathcal{C}_{\alpha}^{\mp}\rangle$ and $a^{\dagger}|\mathcal{C}_{\alpha}^{\pm}\rangle=\alpha p^{\mp 1}|\mathcal{C}_{\alpha}^{\mp}\rangle$, the matrix form of $\mathcal{L}$ in the space spanned by $\{|\mathcal{C}_{\alpha}^{+}\rangle,|\mathcal{C}_{\alpha}^{-}\rangle\}$ can be expressed by a $4\times4$ matrix
\begin{widetext}
\begin{equation}
    \mathcal{L}_{\mathrm{matrix}} =
    \left(
    \begin{array}{cccc}
        -\alpha^2\kappa p^2 & 0 & 0 & \alpha^2\kappa p^{-2} \\
        0 & \alpha^2(-\frac{\kappa}{2}p_2^+ + i\Delta p_2^-)+L_{\phi} & \alpha^2\kappa & -0 \\
        0 & \alpha^2\kappa & \alpha^2(-\frac{\kappa}{2}p_2^+ -i\Delta p_2^-)+L_{\phi} & 0 \\
        \alpha^2\kappa p^{2} & -0 & 0 & -\alpha^2\kappa p^{-2} 
    \end{array}
    \right),
\end{equation}
\end{widetext}
where $p_j^{\pm}=p^{-j}\pm p^j$ and $L_{\phi}=\kappa_{\phi}|\alpha|^4(1-p_4^+/2)$.
Note that,  for conventional qubits, LEPs can be designed independently of basis orthogonality. In our continuous-variable encoding, however, the parameter $p$ indicates the overlap $\langle\alpha | -\alpha \rangle$. When $p\rightarrow1$, this overlap vanishes, $\mathcal{L}_{\mathrm{matrix}}$ reverts to a Hermitian form, precluding any exceptional point.
This reveals a fundamental difference in how state overlap governs non-Hermitian physics.

\begin{figure}[htbp]
	\centering
	\includegraphics[width=2.5in]{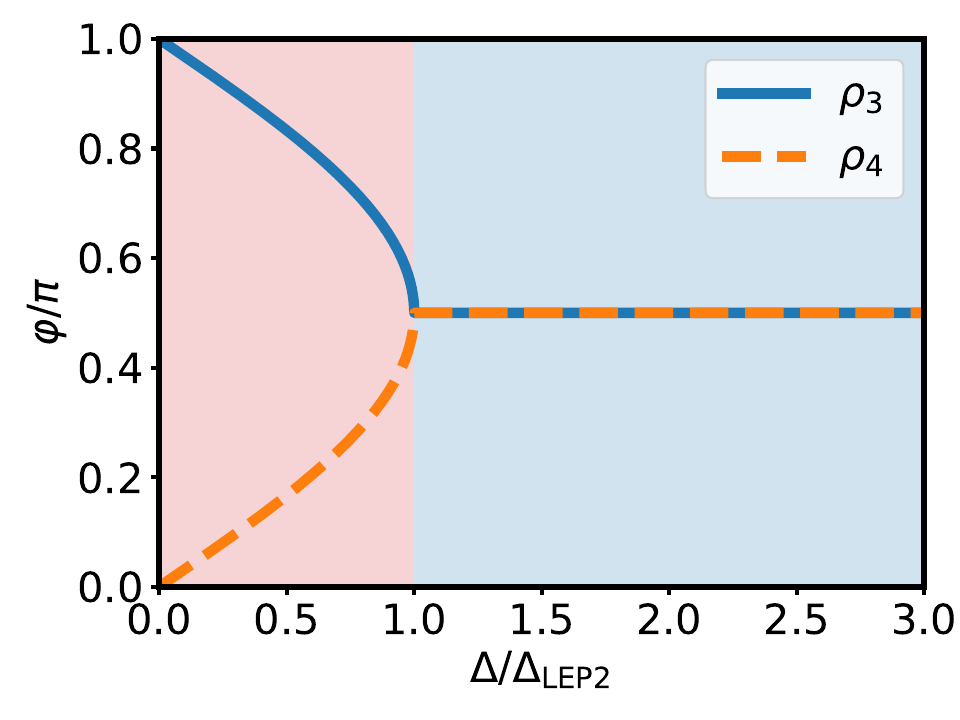}
	\caption{The phase difference ($\varphi$) between the off-diagonal elements of $\rho_3$ (solid blue line) and $\rho_4$ (dashed orange line).}
    \label{Fig5}
\end{figure}

\begin{figure*}[htbp]
	\centering
	\includegraphics[width=7in]{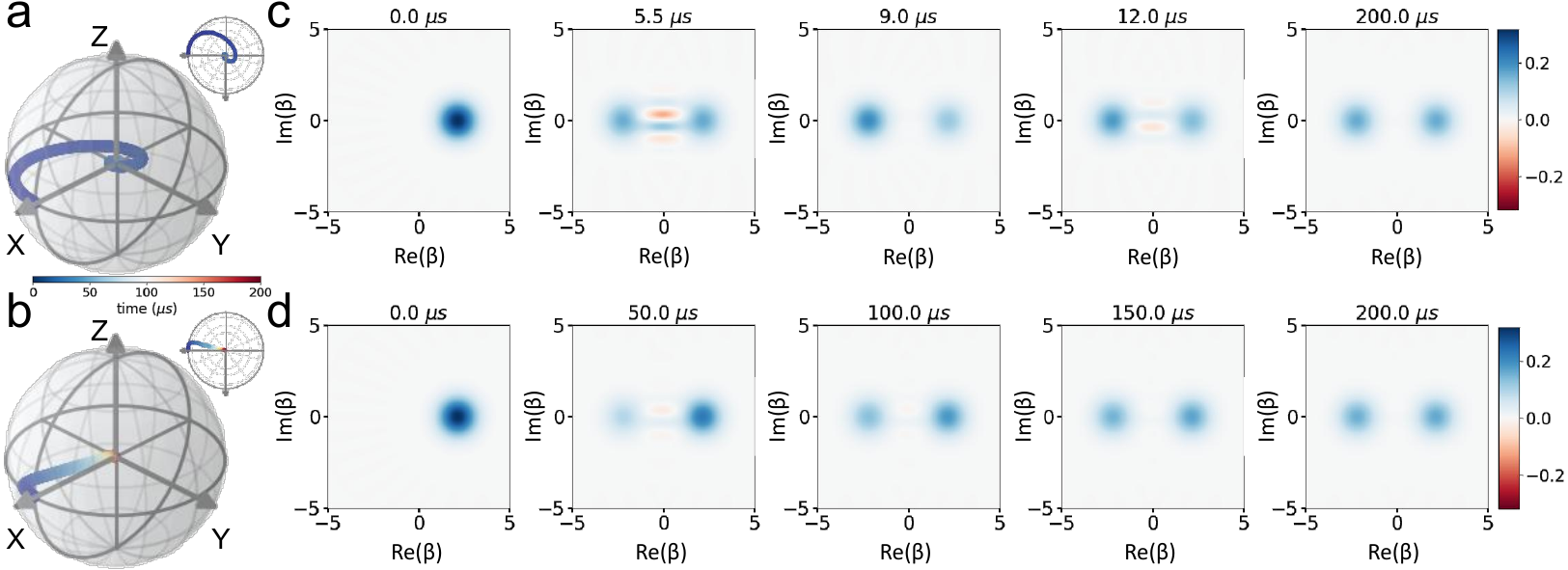}
	\caption{Bloch representation of quantum state evolution and the corresponding Wigner functions.
    Evolution trajectories for (a) $|\Delta|=3\Delta_{\text{LEP2}}$ and (b) $|\Delta|=0.5\Delta_{\text{LEP2}}$. All trajectories are confined to the XY-plane. The inset at the upper right shows a top-down view of the trajectory, with its color corresponding to the evolution time.
    (a) For $|\Delta|>\Delta_{\text{LEP2}}$, the trajectory spirals inward toward the origin (steady state).
    (b) For $|\Delta|<\Delta_{\text{LEP2}}$, the trajectory converges monotonically toward the origin without spiraling.
    Panels (c) and (d) show the corresponding Wigner functions at different evolution times.
    (c) When $|\Delta|>\Delta_{\text{LEP2}}$, interference fringes periodically emerge and vanish before reaching steady state.
    (d) When $|\Delta|<\Delta_{\text{LEP2}}$,  interference fringes appear only transiently before each steady state is achieved.
    The parameters are $\kappa/2\pi=10$ kHz, $K/2\pi=6.7$ MHz, $P/2\pi=15.5$ MHz and $\kappa_{\phi}=0$.}
	\label{Fig3}
\end{figure*}

The Liouvillian $\mathcal{L}_{\mathrm{matrix}}$ yields four eigenvalues
\begin{align}
    E_1 &= 0, \\
    E_2 &= -\kappa |\alpha|^2 p_2^{+}, \\
    E_3 &= \frac{E_2}{2} +L_{\phi}- i |\alpha|^2 \sqrt{ (\Delta p_2^{-})^2-\kappa^2 }, \\
    E_4 &= \frac{E_2}{2} +L_{\phi}+ i |\alpha|^2 \sqrt{(\Delta p_2^{-})^2-\kappa^2}.
\end{align}
The eigenvalue $E_1$ corresponds to the steady state $\rho_{\mathrm{ss}}$, while $E_2$ is strictly real and always associates with the eigenvector $\boldsymbol{\mathrm{Vec}}[\rho_2]=(-1,0,0,1)^{\top}$. The remaining two eigenvalues, $E_3$ and $E_4$, form a complex-conjugate pair, with eigenmatrices satisfying $\rho_3=\rho_4^{\dagger}$.
A second-order exceptional point (LEP2) occurs when the detuning reaches $|\Delta|=\Delta_{\mathrm{LEP2}} \equiv \kappa / p_2^{-}$. At this critical point, $E_3$ and $E_4$ coalesce into a single eigenvalue $E_2/2$ with $\kappa_{\phi}=0$, and their eigenvectors merge into 
$\boldsymbol{\mathrm{Vec}}[\rho_{\mathrm{EP}}] = (0, \; i, \; 1, \; 0)^{\top}$. 
This spectral structure is illustrated in Fig.~\ref{Fig1} for the parameters $\kappa/2\pi=10$ kHz and $\kappa_{\phi}=0$, where the solid green curve demonstrates the exponential scaling of $\Delta_{\text{LEP2}}$ with the cat-state amplitude $\alpha$.

Accordingly, when $\kappa_{\phi}\neq0$, the Liouvillian eigenvalues $E_1$ and $E_2$ remain unchanged, while $E_3$ and $E_4$ exhibit shifts in their real parts by $L_{\phi}$.
Remarkably, the parameters of LEPs remain unchanged.
This result demonstrates that dephasing  modifies exclusively the imaginary components of the corresponding non-Hermitian Hamiltonian spectrum, altering dissipation rates while preserving coherent oscillations—consistent with the phase-noise resilience mechanism for cat states \cite{55}.

As shown in Fig.~\ref{Fig1}, for a fixed value of $\alpha$, the LEP2 separates the Liouvillian spectrum into two distinct regimes. For $|\Delta| < \Delta_{\text{LEP2}}$, all eigenvalues are purely real, whereas a complex spectrum emerges for $|\Delta| > \Delta_{\text{LEP2}}$. The imaginary parts of the eigenvalues set the oscillation periods of the dynamical evolution, while the real parts govern the decay rates. Therefore, the phase transition at the LEP2 can be directly observed from the distinctive dynamical behavior of the state evolution.
The system's dynamical evolution is governed by
\begin{equation}
\rho(t) = c_1 \text{exp}(E_1t)\rho_{ss}+\sum_{i=2}^4 c_i \text{exp}(E_i t) \rho_i,
\label{Liouvillian dynamics}
\end{equation}
where $c_i$ denote expansion coefficients, $E_i$ are the eigenvalues, and $\rho_i$ are the corresponding eigenvectors of $\mathcal{L}_{\text{matrix}}$. Crucially, when $|\Delta|<\Delta_{\text{LEP2}}$, the time evolution of $c_i \text{exp}(E_i t)$ manifests exponential behavior due to the absence of imaginary components in $E_i$, while for $|\Delta|>\Delta_{\text{LEP2}}$, these coefficients display exponentially damped periodic oscillations arising from complex eigenvalue pairs. This dynamical dichotomy defines two distinct phases separated by LEP2s.
The system relaxes to its steady state in the long-time limit.

To observe this phase transition, we initialize the system in the coherent state $|\alpha\rangle$ and apply the two-photon drive with variable duration across different detunings $\Delta$. The simulation parameters are set to $K/2\pi = 6.7$ MHz, $P/2\pi = 15.5$ MHz, and $\kappa/2\pi = 10$ kHz, as reported in Ref.~\cite{48}. From $K$ and $P$, the cat-state amplitude follows as $\alpha = \sqrt{P/K} \approx 1.52$. As demonstrated in \textcolor{black}{Fig.~\ref{Fig2}}, two contrasting dynamical regimes emerge. For detunings exceeding $\Delta_{\text{LEP2}}$, damped oscillations in $\langle X\rangle$($\langle Y\rangle$) signal underdamped dynamics, whereas for $|\Delta| < \Delta_{\text{LEP2}}$, $\langle X\rangle$($\langle Y\rangle$) undergoes monotonic exponential decay characteristic of overdamped behavior. This abrupt transition at $|\Delta| = \Delta_{\text{LEP2}}$ stems directly from Liouvillian eigenvalues and eigenvectors coalescence.

This exceptional phase transition can be further quantified by the phase difference $\varphi$ between the off-diagonal elements of the eigenmatrices $\rho_3$ or $\rho_4$. For an eigenmatrix $\rho_j$ (with $j=3,4$) expressed in the cat-state basis $\{ |\mathcal{C}_\alpha^+\rangle, |\mathcal{C}_\alpha^-\rangle \}$ as
\begin{equation}
    \rho_j =
    \left(
    \begin{array}{ccc}
        \rho_{00} & \rho_{01} \\ \rho_{10} & \rho_{11} 
    \end{array}
    \right),
\end{equation}
the phase difference $\varphi$ is defined as
\begin{equation}
    \varphi \equiv |\arg(\rho_{01}) - \arg(\rho_{10})|,
\end{equation}
where $\arg(\cdot)$ denotes the complex argument. As shown in Fig.~\ref{Fig5}, the variation of $\varphi$ with $\Delta$ provides a clear signature of the phase transition across the LEP. In contrast to previous study that characterized the phase difference between eigenstates of a non-Hermitian Hamiltonian \cite{31}, our work focuses on the phase difference associated with the eigenvectors of the Liouvillian superoperator—a distinct aspect that has not been addressed before.

The dynamical dichotomy is further resolved through phase-space visualization.
The quantum state $\rho$ is fully characterized by its Wigner function
\begin{equation}
W(\beta) = \frac{2}{\pi} \mathrm{Tr} \left[ D(-\beta) \rho D(\beta) e^{i\pi a^{\dagger}a} \right],
\end{equation}
with $D(\beta)$ denoting the displacement operator. This phase-space representation provides complete information about the oscillator's quantum state.
When $|\Delta| > \Delta_{\text{LEP2}}$, quantum-state oscillations between $|\alpha\rangle$ and $|-\alpha\rangle$ periodically generate coherent interference fringes in the Wigner function \textcolor{black}{(Fig. \ref{Fig3}c)}, reflecting non-classical coherence before convergence to $\rho_{ss}$. In contrast, for $|\Delta| < \Delta_{\text{LEP2}}$, the Wigner function shows direct relaxation to $\rho_{ss}$ without oscillatory features \textcolor{black}{(Fig.~\ref{Fig3}d)}. Complementary insights emerge from Bloch sphere trajectories in \textcolor{black}{Fig.~\ref{Fig3}a,b}. The underdamped regime exhibits repeated crossings of the YZ-plane, where a larger detuning $\Delta$ yields more frequent crossings, signifying $\langle X\rangle$ oscillations.
In contrast, the overdamped regime displays radial convergence to the origin without equatorial intersections, confirming the absence of $\langle X\rangle$ switching.

\begin{figure}[htbp]
	\centering
	\includegraphics[width=2.5in]{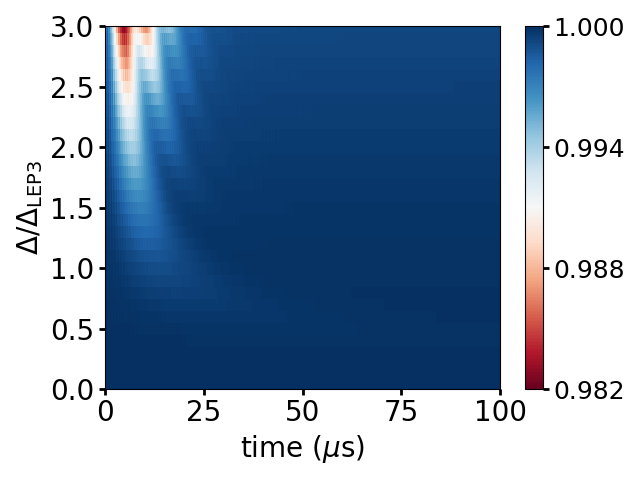}
	\caption{Validation of the Liouvillian dynamics. Color represents the fidelity $F(\rho_{\text{H}},\rho_{\text{L}})$ between states evolved under the Lindblad master equation of Eq.~(\ref{master eq}) and the Liouvillian approximation of Eq.~(\ref{Liouvillian dynamics}).
    Here, we set $K/2\pi = 6.7$ MHz, $P/2\pi = 15.5$ MHz, $\kappa/2\pi = 10$ kHz, and $\kappa_{\phi}=0$.}
    \label{Fig4}
\end{figure}

To quantify how well the system remains confined to the Kerr-cat qubit subspace during evolution, we compare the state $\sigma(t)$ obtained from the effective Liouvillian dynamics of  Eq.~\eqref{Liouvillian dynamics} with the state $\rho(t)$ obtained from the full master equation of Eq.~\eqref{master eq}. Their closeness is measured by the Uhlmann fidelity
$
F(\rho, \sigma) = \left( \operatorname{Tr} \sqrt{ \sqrt{\rho} \, \sigma \, \sqrt{\rho} } \right)^2 .
$
Figure~\ref{Fig4} displays the fidelity as a function of detuning and time for the initial state $|\alpha\rangle$. Over the selected parameter range, the two evolutions show excellent agreement. Furthermore, as the system relaxes toward its steady state, the fidelity asymptotically approaches unity.
The minor deviations stem from the detuning term $\Delta a^\dagger a$, which is not fully captured in the effective Liouvillian description. This term lifts the cat-state degeneracy due to their different photon numbers ($\alpha^2 p^2$ vs. $\alpha^2/p^2$), and can induce leakage out of the encoded subspace.

In summary, we have revealed and characterized a quantum phase transition directly induced by an LEP2 in a single driven-dissipative Kerr-cat qubit. Through analysis of the Liouvillian spectrum and dynamics, we have demonstrated that the transition is marked by a sharp change in dynamical behavior from underdamped oscillations to overdamped relaxations as the detuning crosses the exceptional point. Furthermore, the phase difference between off-diagonal elements of the Liouvillian eigenmatrices emerges as a distinct and previously unreported signature of this exceptional transition. The negativity of the Wigner function serves as a direct hallmark of genuine quantum coherence, a feature unattainable in conventional single-qubit non-Hermitian systems. Our results establish the Kerr-cat qubit as a versatile platform for exploring dissipative quantum criticality and intrinsic non-Hermitian physics, thereby bridging fault-tolerant bosonic encoding with the study of open quantum phase transitions.

This work was supported by the National Natural Science Foundation of China
(Grant Nos. 12505021, 12475015, 12505016), the
Natural Science Foundation of Fujian Province (Grant Nos. 2025J01383, 2025J01465) and the Research Startup Funds of Longyan University (LB2025002).

\end{document}